\input harvmac
%\draftmode
\noblackbox
%-------------------------
% This paper uses harvmac
%-------------------------
 
\font\ticp=cmcsc10
 
\def\Title#1#2{\rightline{#1}\ifx\answ\bigans\nopagenumbers\pageno0\vskip1in
\else\pageno1\vskip.8in\fi \centerline{\titlefont #2}\vskip .5in}

\font\ticp=cmcsc10
\font\ttsmall=cmtt10 at 8pt

%%%%%%%%%%%%%%%%%%%%%%%%%%%%%%%%%%%%%%%%%%%%%%%%%%%%%%%%%%%%%%%
%The following lines are needed to insert the accompanying figures
%in the paper. If you do not have epsf, then comment out the line
% ``\input epsf'', and print the figures separately.
\input epsf
\ifx\epsfbox\UnDeFiNeD\message{(NO epsf.tex, FIGURES WILL BE
IGNORED)}
\def\figin#1{\vskip2in}% blank space instead
\else\message{(FIGURES WILL BE INCLUDED)}\def\figin#1{#1}\fi
\def\ifig#1#2#3{\xdef#1{fig.~\the\figno}
\goodbreak\topinsert\figin{\centerline{#3}}%
\smallskip\centerline{\vbox{\baselineskip12pt
\advance\hsize by -1truein\noindent{\bf Fig.~\the\figno:} #2}}
\bigskip\endinsert\global\advance\figno by1}
%%%%%%%%%%%%%%%%%%%%%%%%%%%%%%%%%%%%%%%%%%%%%%%%%%%%%%%%%%%%%%%%

%
% definitions
%
\def\[{\left [}
\def\]{\right ]}
\def\({\left (}
\def\){\right )}
\def\p{\partial}

\def\S{{\bf S}}
\def\l{\ell}
\def\g{\gamma}

\def\C{{\cal{C}}}

\def\e{\varepsilon}
\def\Om{\Omega}
\def\a{\alpha}
\def\lam{\lambda}
\def\CN{{\cal N}}
\def\CH{{\cal H}}

\def\CI{I}
\def\scri{{\cal I}}

\def\ud{\dot{u}}
\def\vd{\dot{v}}

\def\xid{\dot{x^i}}

\def\rd{\dot{r}}

\def\rd{\dot{r}}

\def\dva{\( {\p \over \p v} \)^a}

\def\xdd{\ddot{x}}

\def\itm{\noindent $\bullet$ \ }
\def\pinf{p_{\infty}}
\def\rinf{r_{\infty}}
\def\vinf{v_{\infty}}

%references

\lref\maro{
D.~Marolf and S.~F.~Ross,
{\it Plane waves: To infinity and beyond!},
[arXiv:hep-th/0208197].}

\lref\geroch{R.~Geroch, E.~H.~Kronheimer, and R.Penrose,
{\it Ideal points in space-time},
Proc.\ Roy.\ Soc.\ Lond.\ A.\ {\bf 327}, 545 (1972).}

\lref\penrose{
R.~Penrose,
``Any spacetime has a planewave as a limit,'' in
{\it Differential geometry and relativity}, pp 271-275,
Reidel, Dordrecht, 1976.}

\lref\penr{
R.~Penrose,
{\it A Remarkable Property Of Plane Waves In General Relativity},
Rev.\ Mod.\ Phys.\  {\bf 37}, 215 (1965).}

\lref\bmn{
D.~Berenstein, J.~M.~Maldacena and H.~Nastase,
{\it Strings in flat space and pp waves from N = 4 super Yang Mills},
JHEP {\bf 0204}, 013 (2002)
[arXiv:hep-th/0202021].}

\lref\host{
G.~T.~Horowitz and A.~R.~Steif,
{\it Space-Time Singularities In String Theory},
Phys.\ Rev.\ Lett.\  {\bf 64}, 260 (1990).}

\lref\bfhp{
M.~Blau, J.~Figueroa-O'Farrill, C.~Hull and G.~Papadopoulos,
{\it A new maximally supersymmetric background of IIB superstring theory},
JHEP {\bf 0201}, 047 (2002)
[arXiv:hep-th/0110242].}

\lref\bena{
D.~Berenstein and H.~Nastase,
{\it On lightcone string field theory from super Yang-Mills and holography},
[arXiv:hep-th/0205048].}

\lref\tseytlin{
A.~A.~Tseytlin,
{\it On limits of superstring in AdS(5) x S**5},
[arXiv:hep-th/0201112].}

\lref\juan{
J. Maldacena, 
{\it The Large N Limit of Superconformal Field Theories and Supergravity},
Adv. Theor. Math. Phys. {\bf 2} (1998) 231, [arXiv:hep-th/9711200].}

\lref\magoo{
O. Aharony, S.S. Gubser, J. Maldacena, H. Ooguri, Y. Oz,
{\it Large N Field Theories, String Theory and Gravity},
Phys. Rept. {\bf 323} (2000) 183, [arXiv:hep-th/9905111].}

\lref\witten{ 
E. Witten, 
{\it Anti De Sitter Space And Holography},
Adv. Theor. Math. Phys. {\bf 2} (1998) 253, 
[arXiv:hep-th/9802150].}

\lref\gkp{
S. Gubser, I. Klebanov, and A. Polyakov, 
{\it Gauge Theory Correlators from Non-Critical String Theory},
Phys. Lett. {\bf B428} (1998) 105,
[arXiv:hep-th/9802109].}

\lref\gueven{
R.~Gueven,
{\it Plane wave limits and T-duality},
Phys.\ Lett.\ B {\bf 482}, 255 (2000)
[arXiv:hep-th/0005061].}

\lref\bfp{
M.~Blau, J.~Figueroa-O'Farrill and G.~Papadopoulos,
{\it Penrose limits, supergravity and brane dynamics},
[arXiv:hep-th/0201081].}

\lref\zv{
L.~A.~Pando Zayas and D.~Vaman,
{\it Strings in RR plane wave background at finite temperature},
[arXiv:hep-th/0208066].}

\lref\gss{
B.~R.~Greene, K.~Schalm and G.~Shiu,
{\it On the Hagedorn behaviour of pp-wave strings and N = 4 SYM 
theory at  finite R-charge density},
[arXiv:hep-th/0208163].}

\lref\sugawara{
Y.~Sugawara,
{\it Thermal Amplitudes in DLCQ Superstrings on PP-Waves},
[arXiv:hep-th/0209145].}

\lref\amkl{
D.~Amati and C.~Klimcik,
{\it Nonperturbative Computation Of The Weyl Anomaly For A Class Of 
Nontrivial Backgrounds},
Phys.\ Lett.\ B {\bf 219}, 443 (1989).}

\lref\gava{
D.~Garfinkle and T.~Vachaspati,
{\it Cosmic String Traveling Waves},
Phys.\ Rev.\ D {\bf 42}, 1960 (1990).}

\lref\garfinkle{
D.~Garfinkle,
{\it Black String Traveling Waves},
Phys.\ Rev.\ D {\bf 46}, 4286 (1992)
[arXiv:gr-qc/9209002].}

\lref\zayasson{
L.~A.~Pando Zayas and J.~Sonnenschein,
{\it On Penrose limits and gauge theories},
JHEP {\bf 0205}, 010 (2002)
[arXiv:hep-th/0202186].
%%CITATION = HEP-TH 0202186;%%
}

\lref\kmr{
N.~Kaloper, R.~C.~Myers and H.~Roussel,
{\it Wavy strings: Black or bright?},
Phys.\ Rev.\ D {\bf 55}, 7625 (1997)
[arXiv:hep-th/9612248].
%%CITATION = HEP-TH 9612248;%%
}

\lref\ct{
M.~Cvetic and A.~A.~Tseytlin,
{\it Solitonic strings and BPS saturated dyonic black holes},
Phys.\ Rev.\ D {\bf 53}, 5619 (1996)
[Erratum-ibid.\ D {\bf 55}, 3907 (1997)]
[arXiv:hep-th/9512031].
%%CITATION = HEP-TH 9512031;%%
}

\lref\cy{
M.~Cvetic and D.~Youm,
{\it General Rotating Five Dimensional Black Holes of Toroidally Compactified 
Heterotic String},
Nucl.\ Phys.\ B {\bf 476}, 118 (1996)
[arXiv:hep-th/9603100].
%%CITATION = HEP-TH 9603100;%%
}

\lref\hormar{
G.~T.~Horowitz and D.~Marolf,
{\it Counting states of black strings with traveling waves},
Phys.\ Rev.\ D {\bf 55}, 835 (1997)
[arXiv:hep-th/9605224].
%%CITATION = HEP-TH 9605224;%%
}

\lref\causal{
V.~E.~Hubeny and M.~Rangamani,
{\it Comments on causal structure of pp-waves},
to appear.
}

\lref\trusso{
J.~G.~Russo and A.~A.~Tseytlin,
{\it A class of exact pp-wave string models with interacting 
light-cone  gauge actions},
JHEP {\bf 0209}, 035 (2002)
[arXiv:hep-th/0208114].
%%CITATION = HEP-TH 0208114;%%
}

\lref\malmaoz{
J.~Maldacena and L.~Maoz,
{\it Strings on pp-waves and massive two dimensional field theories},
[arXiv:hep-th/0207284].
%%CITATION = HEP-TH 0207284;%%
}

%\ChandrasekharJN
\lref\chandra{
S.~Chandrasekhar and B.~C.~Xanthopoulos,
{\it A New Type Of Singularity Created By Colliding Gravitational Waves},
Proc.\ Roy.\ Soc.\ Lond.\ A {\bf 408}, 175 (1986).
%%CITATION = PRSLA,A408,175;%%
}

%\SzekeresUU
\lref\szk{
P.~Szekeres,
{\it Colliding Plane Gravitational Waves},
J.\ Math.\ Phys.\  {\bf 13}, 286 (1972).
%%CITATION = JMAPA,13,286;%%
}

\lref\kpen{
K.~Kahn and R.~Penrose,
Nature {\bf 229}, 185 (1971).
}

%-------------------
% title page
%-------------------
%
\baselineskip 16pt
\Title{\vbox{\baselineskip12pt
\line{\hfil SU-ITP-02/40}
\line{\hfil UCB-PTH-02/48}
\line{\hfil LBNL-51675}
\line{\hfil \tt hep-th/0210234} }}
{\vbox{
{\centerline{No horizons in pp-waves}
}}}
\centerline{\ticp Veronika E. Hubeny$^a$
 and Mukund Rangamani$^{b,c}$ \footnote{}{\ttsmall
 veronika@itp.stanford.edu, mukund@socrates.berkeley.edu}}
\bigskip
\centerline {\it $^a$
Department of Physics, Stanford University, Stanford, CA 94305, USA} 
\centerline{\it $^b$ Department of Physics, University of California,
Berkeley, CA 94720, USA} 
\centerline{\it $^c$ Theoretical Physics Group, LBNL, Berkeley, CA 94720, USA}

\bigskip
\centerline{\bf Abstract}
\bigskip

\noindent
We argue that pp-wave backgrounds can not admit event horizons.
We also comment on pp-wave generalizations which would admit horizons
and show that there exists a black string solution which asymptotes  
to a five dimensional plane wave background.

\Date{October, 2002}
%
%__________________________________________________________________
\newsec{Introduction}

Plane wave\foot{
To avoid any confusion later, let us clarify the terminology
from the very outset:  {\it pp-waves} 
(or ``plane-fronted waves with parallel rays'')
are all spacetimes with
covariantly constant null Killing field; {\it plane waves} are
a subset of these which have in addition an extra ``planar'' 
symmetry along the wavefronts.  These are in fact the spacetimes
that much of the recent literature discussing Penrose limits has
been calling ``pp-waves''.}
 spacetimes have a special importance in theoretical physics.
In general relativity, they form simple solutions to Einstein's
equations with many curious properties.  
They can be thought of as arising from the so-called Penrose limit
\penrose\ of any spacetime, which essentially consists of zooming in
onto any null geodesic in that spacetime.  
Being a subset of pp-waves,
they admit a covariantly constant null Killing field, which in turn 
implies that all curvature invariants vanish.  Nevertheless, they are 
distinct from flat spacetime and their structure is much richer.
Interestingly, as shown by Penrose in \penr, 
these spacetimes are not globally hyperbolic, 
so that there exists no Cauchy hypersurface from which a causal evolution 
would cover the entire spacetime.
This automatically implies that even the causal structure 
of plane waves is different from that of flat spacetime.

pp-wave spacetimes are especially important within
the context of string theory.
This is because they yield exact classical backgrounds for 
string theory, since all curvature invariants, 
and therefore all $\alpha'$ corrections, vanish \refs{\amkl,\host}. 
Hence the pp-wave spacetimes correspond to exact conformal field theories.
Because of this fact, they provide much-needed examples of classical
solutions in string theory, which can in turn be used as toy models 
for studying its structure and properties. 
Plane waves happen to be even simpler, for the action in light cone gauge
is quadratic.

While this fact has been appreciated for some time \refs{\amkl, \host},
only recently have plane waves received significant attention, 
mainly initiated by
the work of Berenstein, Maldacena, and Nastase (BMN) \bmn, based on
the AdS/CFT correspondence \refs{\juan,\witten,\gkp,\magoo}.
These authors
proposed a very interesting solvable model of string theory in 
Ramond-Ramond backgrounds  by taking the Penrose limit of 
$AdS_5 \times \S^5$ spacetime \refs{\bmn, \tseytlin}, the 
holographic dual of $d=4$, $\CN =4$ Super-Yang-Mills theory. 
This limit corresponds to
a particular sector of the gauge theory,
with large dimensions of operators and large charges, but with a finite 
difference between the charges and the dimensions. 
Part of the importance of this result stems from the fact that
since the dual background 
is exactly solvable as a string theory, we can claim to have understood, 
at least in principle, this particular sector of the gauge theory.

The BMN ``correspondence'' has since been examined and generalized 
by many authors and Penrose limits of various supergravity 
solutions have been considered in the recent literature.
One interesting avenue for exploration concerns the addition of black holes
into the plane wave spacetime.  Naively, this might correspond to 
``thermalizing'' the high energy sector of the gauge theory.
While this generalization is rather suggestive and follows in close
analogy with the corresponding developments in the AdS/CFT correspondence,
where adding a large Schwarzschild black hole into AdS corresponds to
thermalizing the gauge theory, no concrete solutions or 
understanding have yet been reached.
Partly, this sector of gauge theory does not yet stand on its own as
a well-defined theory without invoking the limit; but more importantly,
no appropriate black hole solution has yet been found. In the case of the 
maximally supersymmetric homogeneous plane-wave discussed by BMN, exact 
quantization of the light-cone string Hamiltonian helps in the analysis 
of the thermal partition function \refs{\zv,\gss, \sugawara}.

One might naively hope to obtain a black hole
by taking an appropriate Penrose limit of a more general spacetime.
In this paper, we argue that this is {\it not} possible.
In particular, we show that no plane wave can admit event horizons. 
We will in fact make the stronger claim that no  pp-waves can admit 
event horizons.
While the latter may not seem as interesting in the context of the
recent excitement about Penrose limits, it is nevertheless of interest
to string theory, because, as mentioned above, pp-waves are 
exact classical solutions in string theory.  Furthermore, they too
can have interesting duals.

The most obvious way to prove the absence of black holes is 
to examine the global causal structure of
a general pp-wave.
Although it turns out to be a rather formidable task to examine the
causal structure in full generality, in the ensuing paper \causal\
 we will discuss specific examples and
some of the causal properties we expect the general pp-wave to carry.  
It is somewhat simpler to concentrate on just the plane waves.  Indeed,
the causal structure of certain plane waves has recently been studied by
Marolf and Ross \maro, who use 
the approach introduced by Geroch, Kronheimer, and Penrose \geroch,
which is based  on completing the spacetime by ``ideal points'' 
reflecting its causal structure.
Marolf and Ross demonstrate that for homogeneous plane waves, the conformal 
boundary consists of a one-dimensional null line plus two points 
corresponding to future and past infinity.
This result agrees with and generalizes that of \bena, 
who obtain the asymptotic structure
of the BMN plane wave by conformally mapping it
into the Einstein static universe.
In the examples studied thus far it is clear that the entire spacetime
manifold is in the causal past of infinity, thereby precluding the
presence of event horizons.

While the study of pp-wave causal structure has not been completed in full 
generality, 
in this paper, we shall content ourselves with examining the
more limited (but, from string theory point of view,
 perhaps the most interesting) aspect of causal structure, 
which can be addressed in generality.
Specifically, we will ask the question 
{\it can pp-waves admit horizons?}
As revealed above, we will argue that the answer is {\it no}.
This, however, does not mean that there cannot exist black hole/string
solutions
which are asymptotically plane or pp-wave, though they do not respect
the plane or pp-wave symmetries everywhere.
We offer a particular simple example in section 5, 
but work is underway to find more physically interesting solutions.

The outline of this paper is as follows.
In the following short section, we review certain basic aspects of 
plane wave and pp-wave spacetimes, mainly with the view of setting 
up notation, and offer a definition of event horizons and 
asymptopia in spacetimes which are not asymptotically flat. 
We will then motivate an argument for the absence of a horizon 
in the plane wave spacetime, by  showing that any point in the 
spacetime can communicate to arbitrarily large distances using a symmetry
argument. 
Section 4 presents the no-horizon argument for pp-waves.
This is distinct from the arguments presented in section 3, 
but it simultaneously provides an alternate proof that plane waves
can't admit event horizons.
While, from the point of view of constructing interesting black hole
solutions in pp-waves, up to here our results were negative, 
in section 5 we try to remedy this by discussing generalizations
of pp-waves which would admit event horizons.
Finally, we end in section 6 with a more general discussion.

%__________________________________________________________________
\newsec{Terminology and definitions}

To pave the way for arguing why pp-waves cannot admit event horizons,
we first explain what are pp-waves and plane waves 
by writing the corresponding metrics.
We then discuss what it would mean for these metrics to admit black holes,
and offer a criterion for absence of black holes.
In the subsequent sections, we use this criterion to argue that pp-waves
do not admit horizons. Note that we shall be concerned with 
physical spacetimes {\it i.e.}, ones which are solutions to the 
Einstein-Hilbert action with matter content satisfying appropriate 
energy conditions. 

\subsec{Metrics}
To set the notation and re-emphasize terminology, we will write explicitly
three classes of spacetimes, in decreasing generality.
The {\it pp-wave}  spacetimes are defined as spacetimes admitting 
a covariantly constant null Killing field.
The most useful ones\foot{Generically, a  background admitting a 
covariantly constant null Killing field can have non-vanishing 
$g_{u x^i}(u,x^i)$ components of the metric. Also there is no 
requirement that the transverse space be flat; for vacuum solutions we 
could have easily considered Ricci flat transverse metrics. While  
our arguments are expected to hold for these cases, we will 
restrict our discussion to metrics of the form presented in Eq.(2.1).}
can be written as
\eqn\pp{
ds^2 = -2 \, du \, dv - F(u, x^i) \, du^2 + dx^i \, dx^i}
where the vacuum Einstein's equations dictate that $F$ satisfy
the transverse Laplace equation for each $u$.
$F$, however, can be an arbitrary function of $u$. 

{\it Plane wave} spacetimes are those where this harmonic function is in
fact quadratic, $F(u, x^i) = f_{ij}(u) \, x^i x^j$ in \pp,
 so that plane waves can be written as
\eqn\plane{
ds^2 = -2 \, du \, dv - f_{ij}(u) \, x^i x^j \, du^2 + dx^i \, dx^i}
Here, $f_{ij}(u)$ can be any function of $u$, subject to the constraint
that for each $u$,
$f_{ij}$ is symmetric and, for vacuum solutions, traceless.
As suggested by the name, these metrics
have an extra ``plane'' symmetry, which contains the translations
along the wave-fronts in the transverse directions.  This can be seen
explicitly by casting \plane\ into the Rosen form,\foot{
Typically, this metric is not geodesically complete because of 
coordinate singularities, but the Brinkman form \plane\ does 
cover the full spacetime.
The coordinate transformation from one form into the other is given
{\it e.g.}, in \bfp.
For metric of the Brinkman form
 $ds^2 = - 2 \, du \, dv - f(u) \, x^2 \, du^2 + dx^2$, the
coordinate transformation 
$\{ u= U, x=  h(U)\, X , v= V+ {1 \over 2} h(U) \, h'(U) \, X^2 \}$
where $h(U)$ satisfies $h''(U) + f(U) \, h(U) = 0$,
casts this metric into the Rosen form
$ds^2 = -2 \, dU \, dV + h(U)^2 \, dX^2$.
}
\eqn\rosen{
ds^2 = -2 \, dU \, dV + C_{ij}(U) \, dX^i \, dX^j}
The {\it homogeneous plane waves} further specialize \plane\ 
by taking out $f$'s dependence on $u$,
\eqn\homplane{
ds^2 = -2 \, du \, dv - f_{ij} \, x^i x^j \, du^2 + dx^i \, dx^i}
The BMN plane wave metric \bmn, 
found earlier by \bfhp, belongs to this last class, for the special 
case $f_{ij} = \mu^2 \, \delta_{ij}$, and $u \equiv x^+, v \equiv x^-$ 
in their notation.

In all the aforementioned spacetimes we have a covariantly constant 
null Killing vector given as $\xi^a = \( {\p \over \p v }\)^a$.
The fact that this is a null Killing vector is obvious from the metric, 
while its being covariantly constant may be inferred from the vanishing 
of the Christoffel symbols $\Gamma^v_{\ \mu v}$. 

%__________________________________________________________________
\subsec{Event horizons and asymptopia}

Black holes are defined as regions of spacetime inside event 
horizons.  Hence, to show that a particular class of spacetimes
cannot represent black holes, we need to show that these spacetimes
do not admit event horizons.  However, in order to do so, we first
need to specify a suitable definition of an event horizon.

In asymptotically flat spacetimes, an event horizon is defined 
simply as the boundary of the causal past of the future null infinity,
$\CH = \p \CI^-[\scri^+]$, where the past $\CI^-$ of future null 
infinity $\scri^+$ is defined as the union of the pasts of all the 
points $P$ at infinity $\scri^+$, {\it i.e.}, 
$\CI^-[\scri^+] \equiv \cup \CI^-(P) : P \in \scri^+$.
Physically, this just says that an asymptotic observer can't see
inside black holes.
However, as is well-known, when
 the spacetime in question is not asymptotically flat, this simple
definition may not work.  First, the notion of asymptopia may be more
murky.  For instance, as in the case of closed FRW universe, there can be a 
big crunch, so that there does not exist any asymptotic region at all.
Similarly, for some of the presently-studied pp-waves, the ``asymptotic''
region may be singular if $F(u,x^i) \sim x^p$ for some $p > 2$ in 
the pp-wave metric \pp;
or the spacetime may terminate at finite $u$ if $F(u,x^i) \to \infty$
as $u \to u_{\infty} < \infty$.

We will therefore adopt a more universal definition of a black hole,
or rather the absence of black hole, which, instead of requiring that
any point in the spacetime is ``visible'' ({\it i.e.}, causally connected)
to asymptotic observer, merely requires that any point in the spacetime
is visible to an observer who is ``arbitrarily far.''
This last phrase needs a bit more qualification.
One might naively try to define ``arbitrarily far''
 by ``some spatial coordinate getting arbitrarily large,''
but this is too glib
since it is a coordinate-dependent statement.
First of all, if the coordinates don't cover the entire spacetime,
reaching arbitrarily large values of the coordinates would merely 
indicate coming closer to the boundary of our coordinate patch, not the
spacetime.  Also, for all geodesics which don't terminate at a 
singularity, the affine parameter gets arbitrarily large;
but in the case of pp-waves, this will be one of the coordinates, $u$,
which has a spatial component.

The first objection can be bypassed in the case of pp-waves:
the coordinate patch of the metric \pp, with all the coordinates 
ranging form $- \infty$ to $\infty$, does cover the full spacetime. 
The second objection might be mollified by noting that $u$ plays the role
of time rather than a spatial coordinate, but that does not suffice.  
Specifically, 
 there always exists a geodesic, for which $\ud = 0$, 
and $v$ is the affine parameter, so that from any point 
$(u_0, v_0, x^i_0)$ in the spacetime, 
we can causally communicate to $(u_0, v \to \infty, x^i_0)$.\foot{
This may be restated in a more covariant fashion as follows:
For all spacetimes admitting a null Killing field, 
the integral curves along these Killing vectors actually describe 
null geodesics.  To see this,
denote the Killing vector by $\xi^a$.   Then 
$ %\eqn\nullgeodeq{
\xi^a \, \nabla_a \, \xi^b = \xi^a \, \nabla^b \, \xi_a =
{1 \over 2}  \nabla^b (\xi^a \,  \xi_a ) = 0$, %}
where the first equality follows from the definition of Killing vector,
the second from product rule, and the last from $\xi^a$ being null
and thus having a constant norm.
But $\xi^a \, \nabla_a \, \xi^b = 0$ is just the geodesic equation, with 
$\xi^a$ being the tangent vector to the geodesic.}

Since this point is part of $\scri$, the null infinity, 
one might be tempted to argue straight-off that there can't be
event horizons in spacetimes with a null Killing field.
The reason we do not wish to do so is that we don't want to
preclude horizons stretched along the $v$ direction, 
(thus respecting the null symmetry), but separating
say, a region from which no causal curve can reach arbitrarily 
large transverse distance, $x^i$.
(Also, there are counter-examples, such as the black holes
studied by \hormar,\kmr.)

Given the above considerations, we will adopt the following criterion
for absence of black holes in pp-wave spacetimes.

\noindent
{\bf Def:} 
A pp-wave spacetime does not admit an event horizon
iff  from any  point in the spacetime, 
say $(u_0, v_0, x^i_0)$, there exists a causal curve
to some point $(u_1,v_\infty, x^i_\infty)$, where $u_1>u_0$ is arbitrary, 
while $v_\infty, x^i_\infty \rightarrow \infty$. 

\noindent
The important aspect is that not just $v$, but also at least one of the
transverse coordinates, $x^i$, gets arbitrarily big along a causal curve.
We will in fact use a stronger version of this criterion; namely, we 
will require $u_1 = u_0 + \e$, for arbitrarily small $\e > 0$.
This will allow us to use the criterion in greater generality, 
in particular even in the cases where our spacetime terminates at
some finite $u$, {\it i.e.}, $F(u,x^i) \to \infty$
as $u \to u_{\infty} < \infty$.

One more side comment on terminology is in order:  We are using the
term {\it event horizon} in an unconventional (generalized) way, 
as defined above, rather than as something fundamentally related 
to $\scri$.  However, it seems likely that if any part of spacetime is 
visible to an observer who is arbitrarily far, it will also be visible
to an asymptotic observer.  
Also, in the present discussion,  {\it horizon} is used as shorthand for
``event horizon'' as defined above.  It is perhaps worth stressing that
there of course are Rindler horizons in pp-waves, just as there are
Rindler horizons in {\it e.g.} the flat spacetime.  These, however, are rather
trivial, and don't carry any globally special properties.  In particular,
they do not bound a black hole.

%__________________________________________________________________
\newsec{Heuristic motivation for no horizons in plane wave}
 
In the present section we will first give a
heuristic motivation for no horizons in plane wave,
and then try to argue that in plane wave 
spacetimes, causal communication from a given point in the spacetime 
to asymptotically large distances is always possible. 
As discussed above, this 
automatically precludes the presence of horizons.

A heuristic argument for the absence of 
black holes as plane waves is as follows.
As shown originally by Penrose \penrose\ in the context of
classical general relativity, and later extended to 
supergravity by \gueven, a plane wave spacetime 
can be viewed as a Penrose limit of some spacetime.
In this limit, one zooms arbitrarily close to a null geodesic
and reexpands the transverse directions% 
---a procedure analogous to obtaining
a tangent space by zooming in to a point in a manifold---so that
the only nontrivial information which survives is the 
1-dimensional structure 
along the geodesic, parameterized by its affine parameter.
The ``blowing up'' of the transverse directions gives rize to 
the covariantly constant null Killing field mentioned above.
This zooming and reexpanding effectively washes out most of 
the global information contained in the spacetime in all 
directions excepting that along the null geodesic.
In particular, the limit retains local information 
about the spacetime, albeit in a more general fashion than the tangent space,
but at the expense of losing global information such as that pertaining to 
event horizons. 

In the following subsection, we will illustrate this point 
with a specific example, the four dimensional Schwarzschild black hole. 
We will then give a more rigorous argument, essentially
based on the symmetries of the plane wave spacetimes.

%__________________________________________________________________
\subsec{Penrose limits of black hole spacetimes}

We demonstrate that Penrose limits of black
hole spacetimes do not retain information about the event horizon
({\it cf.} 
\zayasson, for considerations of Penrose limits in AdS-Schwarzschild 
spacetimes).
While this is somewhat obvious from the preceding discussion,
it may nevertheless serve as an intuition-building exercise.
Consider, for instance, the asymptotically-flat Schwarzschild 
black hole. The causal structure is as given by the Penrose diagram 
of Fig.1. Since the Penrose limit requires us to consider the neighbourhood of 
null geodesics, let us see what sorts of 
null geodesics are allowed in the spacetime. 

\ifig\figSchw{Penrose diagram of the Schwarzschild black hole.
The curves $g1$, $g2$, and $g3$ describe different null geodesics 
which we might consider for taking the corresponding Penrose limit.}
{\epsfxsize=8cm \epsfysize=4cm \epsfbox{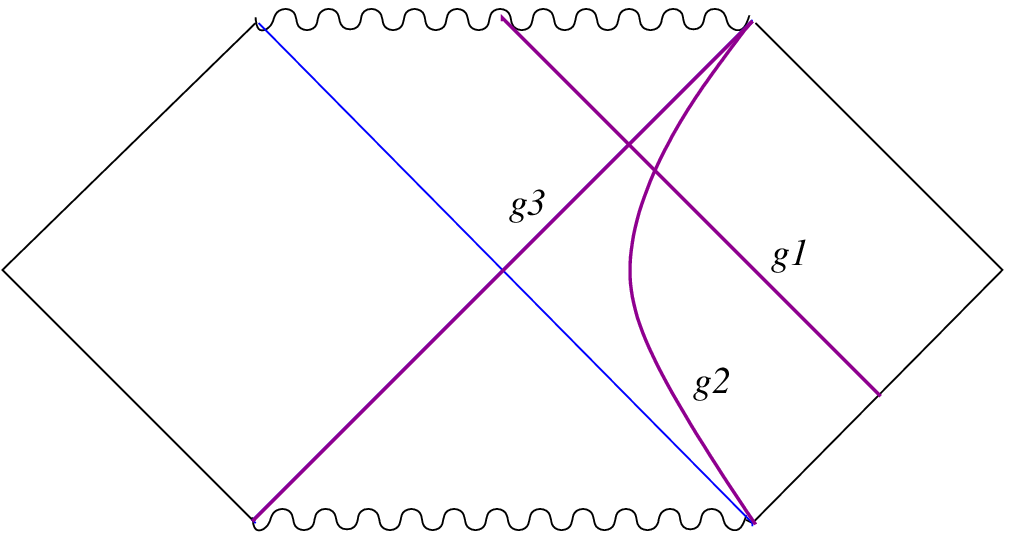}}

First of all, it is clear that there are radially infalling geodesics, 
such as $g1$ of Fig.1,  which 
describe the trajectory of a photon falling into the black hole. 
There can also be ones with angular momentum (which in the $r-t$ plane
of the Penrose diagram would appear timelike), which likewise fall into 
the singularity.
These geodesics intersect the horizon at a single point and terminate at 
finite value of affine parameter upon hitting the singularity. 
The resulting plane wave spacetime will have a singularity reflecting this, 
and in fact the spacelike singularity of the Schwarzschild black hole 
will be converted to a ``cosmological'' null singularity. 
From this 
construction it is clear that the resulting spacetime will not have 
a horizon since we keep only a small region close to the 
point on the horizon where the geodesic intersects the same.  In particular,
this geodesic would be completely insensitive to, for instance,
 a null shell which
might fall into the black hole later, thus shifting the position of the
event horizon of the original spacetime.

A second class of null geodesics are those which are carrying some angular 
momentum, but staying put at constant values of the radial coordinate,
such as the curve labeled by $g2$ in Fig.1. 
This physically corresponds to photon orbits in the black hole spacetime. 
For the four dimensional Schwarzschild black hole this happens to be 
at $r = 3 M$.
However, the neighbourhood of this region is completely 
smooth and the resulting Penrose limit is just the flat space. 

The last interesting geodesic which may be considered is one which is 
sitting at the horizon {\it i.e.}, 
$r=2 M$, and some constant angle $\Om=\Om_0$; this is labeled by $g3$ in 
Fig.1.
This would be the geodesic just skimming the horizon and one would be most 
tempted to consider this as the 
one which can lead to an interesting spacetime in 
the Penrose limit. This geodesic also leads to a flat space in the Penrose 
limit for reasons similar to the previous case.

To summarize, Penrose limits of black hole spacetimes are incapable of 
retaining the global structure of the event horizon.
We note in passing that we are here strictly considering Penrose limits 
of given spacetimes which has a well-defined algorithmic prescription. 
It is less clear whether there exist other limiting
procedures (probably double/mutiple scaling limits), wherein we start 
with a spacetime with a horizon and 
end up with a resulting simple spacetime (some analog of plane wave), 
whilst retaining interesting information about the global causal structure. 

%__________________________________________________________________
\subsec{geodesics and symmetries in general plane waves}

Let us consider the general plane wave metric \plane.
As can be easily checked, the null geodesics are given by
\eqn\geodx{
\xdd^i + \sum_j f_{ij}(u) \, x^j  = 0}
and the integral constraint 
\eqn\veq{
v = {1 \over 2} \, \sum_i x^i \, \xid + v_0}
where $v_0$ is an arbitrary integration constant which is fixed by the 
initial conditions.
The null Killing field $\dva$ implies that $\ud$ is a constant of motion,
so that we can take $u$ to be the affine parameter along the geodesic, 
and the derivative $\ \dot{} \equiv {d \over du}$.

Let us now make the following simple observation:
Under the constant rescaling 
$x^i \to \lambda \, x^i$, $v \to \lambda^2 \, v$, 
the metric $ds^2 \to \lambda^2 \, ds^2$ remains physically the
same (only the ``units'' get rescaled).
Therefore any geodesic 
\eqn\resc{
\{u, v(u), x^i(u)\} \longrightarrow 
\{u,  \bar{v}(u), \bar{x}^i(u)\} \equiv
\{u, \lambda^2 \, v(u), \lambda \, x^i(u)\}}
remains the same under this rescaling.  
Note that this is exactly what we would expect from the 
geodesic equation: since \geodx\ is linear in $x^i(u)$, we are
free to rescale $x^i$; and from \veq, $v$ is rescaled as $(x^i)^2$.

This rescaling freedom suggests that if a geodesic can make
it to some distance $x^i$, it can make it to arbitrarily large $x^i$, so
there couldn't be a horizon at any finite value of $x^i$.
Note that this is already obvious for $v$, since $\dva$ is
a Killing field, so no value of $v$ can be physically distinguished
from any other.

While these arguments were mostly motivational, 
in the next section we shall present  more rigorous proof
of nonexistence of event horizon in pp-waves, which include
plane waves as a subset.

%__________________________________________________________________
\newsec{No horizons in pp-waves}

In the above, we have shown that there can be no horizons in a
generic plane wave.
Let us now ask a more general question, namely, 
{\it can there be horizons in a pp-wave?}

We shall be working with the pp-wave metric \pp, but since 
we will be interested in curves which reach large transverse
directions $x_i$, it is more convenient to rewrite \pp\ in spherical 
coordinates $\{ r,\Om \}$, where only $r$ can get large.
The $d$-dimensional pp-wave metric then becomes
\eqn\pps{
ds^2 = -2 \, du \, dv - F(u,r,\Om) \, du^2 
                      + dr^2 + r^2 \, d\Om_{d-3}^2}
As indicated above, in order to demonstrate the absence of horizons, 
it suffices to show that from {\it any } point 
$p_0=(u_0,v_0, r_0, \Omega_0)$ of  the spacetime,
there exists a causal curve $\C$ which reaches
arbitrarily large values of $r$ and $v$ in arbitrarily small
$\Delta u$. 
Below, we will first try to construct such curves
explicitly, and then offer a more elegant proof.

As apparent from \pps, any causal curve $\C(\lambda)$ must satisfy
\eqn\causalppu{
-2 \, \ud \, \vd -F(u,r,\Om) \, \ud^2 + \dot{r}^2 + r^2 \, \dot{\Om}^2
\le 0}
where $\dot{} \equiv {d \over d\lambda}$.
Then $\C_0(\lambda)
 \equiv \{ u(\lambda)\!=\!u_0, v(\lambda)\!=\!\lambda + v_0,
r(\lambda)\!=\!r_0, \Om(\lambda)\!=\!\Om_0 \}$ is a causal 
(and in fact null)
curve which reaches arbitrarily large values of $v$.  However, 
it stays at constant $r$.  Since all curves with $\ud = 0$ are
simply related to $\C_0$, let us now consider curves with $\ud > 0$.
This will be necessary in order for the curve to reach arbitrarily
large $r$.
We can then rewrite the causal relation \causalppu\ simply as
\eqn\causalpp{
-2 \, \vd -F(u,r,\Om) + \dot{r}^2 + r^2 \, \dot{\Om}^2 \le 0}
where now $\dot{} \equiv {d \over du}$.
Let $\C(u) = \{ v(u),r(u),\Om(u) \}$ be a curve such that
\eqn\pnullc{\eqalign{
\dot{\Omega} & = 0 \;\;\; \Rightarrow \;\;\; \Omega = \Omega_0 \cr
\dot{r}^2 &=  \a \, F(u,r, \Om) \cr
2 \, \dot{v}  & = (\a - 1) \, F(u,r, \Om) 
= (1 - {1 \over \a}) \, \dot{r}^2
}}
where $\a$ is some constant, to be chosen later, % $\a<1$, 
and the initial conditions are given by $p_0$. 
Now, if %$\ud$ is very tiny, such that 
$\Delta u \ll 1$ along the full curve,
we can approximate the spacetime region through which such a
curve propagates by replacing $F(u,r, \Om)$ with $F(u_0,r, \Om)$.
Furthermore, since our curve stays at fixed $\Om = \Om_0$, we
can ``freeze'' that dependence in $F$ as well.
Thus, letting $f(r) \equiv  F(u_0,r, \Omega_0)$, our
curve $\C$ is arbitrarily well approximated by 
\eqn\capprox{\eqalign{
\dot{\Omega} & = 0 \;\;\; \Rightarrow \;\;\; \Omega = \Omega_0 \cr
\dot{r}^2 &= \a \, f(r)  \cr
2 \, \dot{v}  & = (\a - 1) \, f(r)
= (1 - {1 \over \a}) \, \dot{r}^2
}}
Since $\C$ is constructed so as to satisfy \causalpp, it is clearly 
a causal curve.  Hence we only need to show that $\C$ 
 exists and can reach
arbitrarily large $r$ and $v$ for some choice of $\a$.
But \capprox\ is a first order system, which we can just integrate forward; 
the requirement for the existence of the solution is that
both $ \dot{r}^2$ and $\vd$ remain non-negative.
If $|f(r)|$  
is bounded from below,
we can pick $\a$ such that the radial 
velocity of the curve is always positive, and $\C$ reaches arbitrarily 
large $r$.

Construction of a suitable curve is going to be more problematic in 
regions where $f(r)$ changes sign. To specify the curve completely 
 we first pick the sign of $\alpha$ depending on the sign of $f(r)$
near $r = r_0$. Now suppose $f(r)$ passes through zero
 for some $r=r_1 \, > \, r_0$.
At $r_1$ we also flip the sign of $\alpha$ and continue with the construction 
of the causal curve. In other words, in each interval in $r$ where
$f(r)$ doesn't change sign, we can solve the equation
$\dot{r}^2 = \a \, | f(r) |$, and then patch the outgoing pieces together.
 Of course, such a curve will be 
continuous but will have discontinuities in its second derivatives. 

While this explicitly constructs a causal curve, the technique used to 
construct the same is not quite elegant. 
In particular, it requires us to separate the spacetime into various regions 
depending on the sign of $f(r)$, construct a causal curve which reaches
maximal $r$ in each specific region, and then patch the pieces together.
To avoid this cumbersomeness, we will now present an
alternate proof which is more universal.

\itm
Pick any point  $p_0=(u_0,v_0, r_0, \Omega_0)$ in  the spacetime, 
and any $\rinf$, $\vinf$ 
(which will represent the arbitrarily large distances that we want
to reach by a causal curve).
To prove the absence of horizons,
we want to show that there exists a point $\pinf=(u_0+\e,v,\rinf,\Om)$,
with some $\Om$, $v \ge \vinf$, and $\e$ arbitrarily small, %\foot{
which lies on a causal curve $\g$ from $p_0$.

\itm
Pick a constant, $F_0$, such that 
$F(u,r,\Om) \ge F_0 \ \ \forall \ \ r \in (r_0, \rinf)$.
This will clearly be possible if $F(u,r,\Om)$ is not singular
in this region, but we can generalize the proof for singularities 
as well.  We will discuss the existence of $F_0$ below; 
but for the moment, we will assume that $F_0$ does exist.

\itm
Now, consider the fiducial metric, 
\eqn\ppfid{
ds_0^2 = -2 \, du \, dv - F_0 \, du^2 
                      + dr^2 + r^2 \, d\Om_{d-3}^2}
We want to claim that any curve $\g$ which is causal in $ds_0^2$
is also causal in $ds^2$.  But this is easily shown:\foot{
We will now revert back to $\lambda$ parameterizing our curve,
so $\dot{} \equiv {d \over d\lambda}$.}
Any curve $\g$ which is causal in $ds_0^2$ must satisfy
$-2 \, \ud \, \vd -F_0 \, \ud^2 + \dot{r}^2 + r^2 \, \dot{\Om}^2
\le 0$, but since $F(u,r,\Om) \ge F_0$, this automatically
implies that \causalpp\ is also satisfied.
This means that if we can find a curve $\g$ from $p_0$ to 
$\pinf$ which is causal in $ds_0^2$, we are done.

\itm
But that is also easy.
To find $\g(\lam)$ such that
\eqn\conds{\eqalign{
 \g(0) &= (u_0,v_0, r_0, \Omega_0)   \cr
 \g(1) &= (u_0+\e ,v \ge \vinf, \rinf , \Omega_0)   \cr
{\rm and } \ \ \ \ 0 &\ge 
-2 \, \ud \, \vd - F_0 \, \ud^2 
                      + \rd^2 + r^2 \, \dot{\Om}^2}}
let $\g(\lam)$ be given by {\it e.g.} 
\eqn\exgamma{\eqalign{
 u(\lam) &= u_0 + \lam \, \e   \cr
 v(\lam) &= v_0 + \lam \, \Delta v  \cr
 r(\lam) &= r_0 + \lam^2 \, \Delta r  \cr
 \Om(\lam) &= \Om_0   }}
with $\e$ and $\Delta r \equiv \rinf - r_0$ fixed, and $\Delta v $ to be
chosen so as to satisfy the causality condition of \conds, namely
$2 \e \, ( \Delta v + F_0 \, \e) \ge 4 \Delta r^2 \, \lam^2$ for 
$0 \le \lam \le 1$.
Let 
\eqn\Deltav{
 \Delta v = {\rm Max} \{ \vinf - v_0, 
                    {2 \Delta r^2 \over \e} + |F_0| \, \e \} }
Then $\g(\lam)$ satisfies all the requirements of \conds;
in particular, it is causal in $ds_0^2$.  
As argued above, this also means that it is causal in the pp-wave
spacetime $ds^2$.

\itm
In fact, this is easy to see, since $ds_0^2$ is just the flat 
spacetime. 
Explicitly, if $F_0 > 0$, consider the coordinate transformation
$u = {1 \over \sqrt{F_0}} \, (t-z)$ and $v= \sqrt{F_0} \, z$; 
while if $F_0 < 0$, consider the coordinate transformation
$u = {1 \over \sqrt{-F_0}} \, (t-z)$ and $v= \sqrt{-F_0} \, t$.
In both cases, the metric \ppfid\ becomes
$ds_0^2 = -dt^2 + dz^2 + dr^2 + r^2 \, d\Om_{d-3}^2$, which is clearly 
the $d$-dimensional Minkowski spacetime.
But we know that Minkowski spacetime has no horizons, so that 
from any point, there exists a causal curve which can attain 
arbitrarily large values of $z$ and $r$.

Thus, we have found a causal curve starting from an arbitrary point $p_0$
of the pp-wave spacetime and attaining arbitrarily large values 
of the coordinates.  Hence no point $p_0$ can be inside an
event horizon, so that there can't be black holes.
The only step which still needs to be discussed is the existence of
$F_0$, to which we turn next.

%__________________________________________________________________
%\subsec{vacuum pp-waves}
Let us first concentrate on the class of pp-waves which are
solutions to vacuum Einstein's equations.
Since the Einstein tensor is given by
$G_{uu} = {1 \over 2} \nabla_T^2 F$, where $\nabla_T^2$ is the
transverse Laplacian,
 $F(u,r,\Om)$ of \pps\
must satisfy the transverse Laplace equation, $\nabla_T^2 F = 0$.
This is a very remarkable result, since this implies that,
due to the linearity of Laplace equation, 
we may superpose the solutions.
In particular, we can decompose $F$ in terms of the generalized
$(d-3)$-dimensional spherical harmonics $Y_L(\Om)$, 
where $L \equiv \{ \l,m,... \}$:
\eqn\F{
F(u,r,\Om) = \sum_L \{ f_L^+(u) \, r^{\l} \, Y_L(\Om)
          +  f_L^-(u) \, r^{-(d-4+\l)} \, Y_L(\Om) \} }

Therefore along our curve $\g$, this becomes 
\eqn\Fapprox{
F(u_0,r,\Om_0) 
= \sum_{\l} \{ f_{\l}^+ \, r^{\l}  +  f_{\l}^- \, r^{-(d-4+\l)} \} }
We see that there can be singularities at $r=0$ and $r= \infty$.
Now, all ``singularities'' are by definition excluded from our spacetime,
in the sense that all points which are part of the physical spacetime 
must be nonsingular.  In particular all starting points $p_0$ 
must be of that kind.  Thus, if there is a singularity at $r=0$,
we must chose $r_0>0$.  Similarly, we must chose $\rinf < \infty$.
But then in the region of interest, $r_0 \le r \le \rinf$, $F$ is
bounded.  This shows that $F_0$ must always exist for vacuum pp-waves.

What about non-vacuum solutions?  This is much more complicated 
to analyse,
since $F(u,r,\Om)$ can in principle be anything as long as we have the 
appropriate matter content.  
{\it A-priori}, it could obstruct the proposed path of $\g$ by a singularity.
For the case of pp-wave solutions that lead to integrable sigma models 
in light-cone quantization of the world-sheet superstring theory \malmaoz,
\trusso, it is possible to see that we can construct causal curves reaching 
the asymptotic regions of the spacetime. 

%_______________________________________________________________________
\newsec{Generalizations admitting horizons}

Above, we have seen that for pp-wave spacetimes, namely those admitting
a covariantly constant null Killing field,
there can't be any event horizons.
The existence of this null Killing field played an important role in
this observation; in fact, 
the integral curves of this null Killing vector define null geodesics,
so that as we have argued at the beginning, 
from any point in such a spacetime we can ``communicate out to infinity''
at $v = \infty$.
However, as we also cautioned, this does not automatically guarantee
the absence of horizons: without horizons, we should be able to 
reach infinity in all directions. What is most remarkable is the 
fact that one is able to communicate causally also in the transverse 
directions out to large distances.

We therefore want to ask, how many of the properties of pp-waves 
do we need to relax, in order for the existence of horizons to become
possible. We would like to claim that we {\it can} find black hole solutions 
 admitting a null Killing field, which however is not covariantly constant. 
While we have no real evidence for this claim, it is easy to see 
{ \it post facto}
that the relaxation of the covariant constancy requirement does lead to 
spacetimes with an event horizon. In fact, such spacetimes already 
exist in literature. The simplest such example is the 
case of traveling waves on a five-dimensional black string as 
discussed in \kmr\ ({\it cf.}, \ct, \cy, and \hormar\ 
for additional discussions on related issues). 

The solution studied in \kmr\ is a solution to the low energy 
effective action for the heterotic string in five dimensions. 
The metric and the dilaton for the solution are given by
\eqn\fdbh{\eqalign{
ds_{str}^2 &= {2 \over h(r)}  \, du\, dv + {f(r) \over h(r)}
 \, du^2 + k(r) \, l(r) \(
dr^2 + r^2 d\Omega_2^2 \) \cr
e^{4\phi} & = {k(r) \, l(r) \over h^2(r)} 
}}
The metric has been written in the string frame and the Einstein frame 
metric is given as $G^{(E)}_{\mu \nu} = g^{(str)}_{\mu \nu} e^{-4\Phi/3}$.
The functions appearing in the metric are given by
\eqn\fhklfns{\eqalign{
f(r) & = 1 +{Q_1 \over r}  \;\;\;\;\;\;\;\;\;\; h(r) = 1 + {Q_2 \over r} \cr
k(r) & = 1 + {P_1 \over r} \;\;\;\;\;\;\;\;\;\; l(r) = 1 + {P_2 \over r} \cr
}}
There are other fields which need to be turned on for the above metric 
to solve the equations of motion and we refer the reader to the 
original source \kmr\ for explicit expressions of the same.
By a judicious choice of the charges we can even set the dilaton to be 
constant; setting $Q_2 = P_1 = P_2$ will suffice for the same.
 
These spacetimes have an event horizon at $r=0$, which has a finite 
area. As $r \rightarrow 0$ we see that the radius of the two-sphere 
takes the constant value $\sqrt{ P_1 P_2}$. So we have a finite area and 
therefore a solution with finite entropy.
While the coordinates in which the above metric is 
written degenerate near the horizon, it is possible to find a set 
of regular coordinates \kmr. The solution is asymptotically flat since the 
functions appearing in the metric go over to unity for large values of the 
radial coordinate.

One can imagine recovering a spacetime with a covarinatly constant 
null Killing vector\foot{We would like to thank Gary Horowitz for 
bringing this to our attention.} from the solution given in \fdbh, by setting 
$h(r) =1$, which requires the choice $Q_2 =0$. This gives a string frame 
metric with a regular horizon at $r =0$ and $\( {\p \over \p v}\)^a$ is a 
covariantly constant null Killing field, seemingly violating the 
claims we have made hitherto. However, this is illusory. 
In the physical spacetime {\it i.e.},  in the Einstein frame metric, 
we see that $\( {\p \over \p v}\)^a$ is no longer covariantly constant.
In addition $r =0$ is a singular point in the spacetime, for the curvature 
invariants blow up there. It is interesting, however, that while there 
is no contradiction in the physical spacetime, there apparently is a 
violation of our claims in the string frame metric.

One can use the above charged black string solution and generate a solution
which is asymptotically plane-wave. The idea is to use the Garfinkle-Vachaspati
construction \gava, \garfinkle\ to make the spacetime asymptotically 
plane-wave. As explained in \kmr, this is possible for spacetimes which 
admit a null Killing vector which is hypersurface orthogonal and also 
show that this procedure leads to spacetimes which have the same 
set of curvature invariants as the original spacetime. 

For the particular case of the asymptotically flat black string, 
this construction implies that we can add a term 
${\Psi(u,r,\Omega) \over h(r)} \, du^2$ 
to the metric appearing in \fdbh; the resulting metric is a solution 
to Einstein's equations with all other fields unaltered, so long as 
$\Psi(u,r,\Omega)$ is a harmonic function in the transverse space. 
In particular, we can have
\eqn\harmonic{
\Psi(u,r,\Om) = \sum_{\l} \{ \psi_{\l}^+(u) \, r^{\l} \, Y_{\l m}(\Om)
          +  \psi_{\l}^-(u) \, r^{-(1+\l)} \, Y_{\l m}(\Om) \}, 
}
with arbitrary functions $\psi_{\l}^{\pm}(u)$. All spacetimes with 
$\Psi(u,r,\Om) \sim r^{\l}$ for $\l > 2$ suffer from singular behaviour at 
$r = \infty$, while those with $\Psi(u,r,\Om) \sim r^{-(\l+1)}$ with 
$\l > 0$ are singular at $r =0$. By singular we mean that there 
are divergent tidal forces on finite-sized observers.
Addition of $\psi_0^+(u)$ and 
$\psi_1^+(u) \, r \, Y_{1m}(\Om)$ 
lead to spacetimes which are diffeomorphic to 
the original, and the monopole solution leads to a regular spacetime, as 
was demonstrated by \kmr. These cases are the most uninteresting ones 
as far as constructing a black hole spacetime which is asymptotically 
plane wave. The interesting case therefore is the case when
$\Psi_{plane}(u,r,\Om) = \psi^+_2(u) \, r^2 \, Y_2(\Om) = 
\psi^+_2(u) \, (2 z^2 - x^2 - y^2)$, 
reverting back to cartesian coordinates. Now it is 
clear  that $ds_{str}^2 + {\Psi_{plane}(u,r,\Om) \over h(r)} du^2$ is an
asymptotically plane wave spacetime, whilst retaining the regular 
black hole horizon at $r=0$. 
These statements of course remain true in the Einstein frame as well.
The string frame metric for the solution is then given as (setting 
$\psi^+_2(u) = 1$),
\eqn\fdbhplane{\eqalign{
ds_{str}^2 & = 
{2 \over h(r) }  \, du\, dv + {f(r) + 
 r^2 \( 3 \, \cos^2 \theta -1 \)\over h(r) }
 \, du^2 + \(k(r) l(r) \)^2 \, \(
dr^2 + r^2 \, d\Omega_2^2 \) \cr
e^{4\phi} & = {k(r) \, l(r) \over h^2(r)}, 
}}
where the functions $f(r)$, $h(r)$, $k(r)$, $l(r)$ are given in \fhklfns.
Once again the Einstein frame metric is 
$G^{(E)}_{\mu \nu} = g^{(str)}_{\mu \nu} e^{-4\Phi/3}$.
The essential trick in constructing the same 
is that close to the origin, the plane wave is identical to flat space and 
so given a spacetime which has a horizon at $r=0$, we can make it 
asymptotically plane wave whilst keeping the horizon.

The above construction provides an example of a charged black string which 
is asymptotically of the plane wave form. This naturally begs the question 
whether there isn't a neutral black string which is asymptotically a plane 
wave. We cannot ``uncharge''
 the above solution to get a neutral solution, as then 
the horizon shrinks to zero size; setting any of the charges to be zero 
causes the horizon to shrink toward the singularity.  However, we should 
be able to 
take two such solutions with opposite charges and collide them. Since each 
such black string has a finite horizon area, the area theorem will tell us 
that the resulting solution ought to have a horizon of finite area. 
Colliding two such solutions shouldn't change the asymptotics and hence 
we should have a spacetime which is a neutral black string with a finite 
entropy and asymptoting to a plane-wave spacetime.

%__________________________________________________________________
\newsec{Discussion}

In the first four sections of this paper,
we have established that pp-waves cannot admit event horizons.
While this is easily motivated for plane waves, 
we have provided an alternate argument for the more general pp-waves,
which in particular applies to plane waves. For the particular case of 
plane waves one can argue for the absence of horizons in a  
more rigorous fashion following the analysis of the causal structure of these 
spacetimes in \maro. We shall present similar arguments for pp-waves 
in a future work \causal.
 
The reason for considering pp-waves  rather than just
plane waves is that for both classes, all the curvature invariants
vanish, so that they represent exact classical solutions to string 
theory. Also, given a plane wave background, one simple deformation of the 
same is to convert it into a pp-wave background. This follows trivially from 
the fact that the Einstein's equations for the metric ansatz in \pp\ are 
linear, enabling superposition of the solutions. In fact, this is the 
simplest example of the Garfinkle-Vachaspati construction \gava,\garfinkle. 
In a sense, these deformations are similar to exactly marginal deformations in 
the world-sheet theory, though they are non-normalizable. 
It is then perhaps somewhat disappointing that these classes of exact
classical solutions cannot admit horizons.

However, this does not mean that these solutions cannot be modified 
to include horizons.  After all, on physical grounds, one would expect that
if one puts some matter into a plane wave which respects the necessary
symmetries, this matter may nevertheless be Jeans-unstable to collapsing.
Naively, one may then expect to obtain a black hole.  This of course
does not conflict with our previous conclusions, since such black holes
would break the original symmetries.  In particular, the geometry
would no longer support a covariantly constant null Killing field.
It would be very interesting to study this Jeans instability, and to
follow the evolution dynamically, but that is beyond the scope of the
present paper.

Instead, in the previous section,
we have explored only a mild relaxation of the pp-wave symmetries,
namely, keeping the null Killing field, and dropping only
the requirement that it be covariantly-constant.
Hence, this is not a pp-wave, and the curvature invariants do not vanish.
We have presented an explicit solution of a black string with a horizon 
which is asymptotically a plane wave, in five-dimensions. It would be 
very interesting to study solutions which do not carry any charges. One 
strategy as mentioned earlier would be try to collide two oppositely 
charged asymptotically plane wave black strings. The collision of 
plane waves is a problem that has been discussed hitherto in 
\szk, \kpen, \chandra. Perhaps it would be possible to extend these 
discussions to the charged black string discussed above.

%__________________________________________________________________

\vskip 1cm

\centerline{\bf Acknowledgements}
It is a great pleasure to thank Gary Horowitz, Nemanja Kaloper, and Don Marolf 
for illuminating discussions. We in addition would like to thank 
Gary Horowitz and Don Marolf for their comments on the manuscript. 
We gratefully acknowledge the hospitality of Aspen Center for Physics,
where this project was initiated.
VH was supported by NSF Grant PHY-9870115, while MR acknowledges support 
from the Berkeley Center for Theoretical Physics and also partial support 
from the DOE grant DE-AC03-76SF00098 and the NSF grant PHY-0098840.

%
% ====================================================
%

\listrefs
\end